\def\be{\begin{equation}}
\def\ee{\end{equation}}
\def\bea{\begin{eqnarray}}
\def\eea{\end{eqnarray}}
\def\bse{\begin{subequations}}
\def\ese{\end{subequations}}

\documentclass[prl,twocolumn,showpacs,amsmath,amssymb,preprintnumbers]{revtex4}
\usepackage{graphicx}
\usepackage{dcolumn}
\usepackage{bm}
\begin{document}
\title{Criticality in inhomogeneous magnetic systems: Application to quantum
       ferromagnets}


\author{D. Belitz$^{1,2}$, T.R. Kirkpatrick$^{3}$, and Ronojoy Saha$^{3,1}$}
\affiliation{$^{1}$Department of Physics and Materials Science Institute,
              University of Oregon, Eugene, OR 97403\\
             $^{2}$ Institute for Theoretical Science, University of Oregon, Eugene,
                 OR 97403\\
             $^{3}$ Institute for Physical Science and Technology, and
                    Department of Physics, University of Maryland,
                    College Park, MD 20742
          }
\date{\today}
\begin{abstract}
We consider a $\phi^4$-theory with a position-dependent distance from the
critical point. One realization of this model is a classical ferromagnet
subject to non-uniform mechanical stress. We find a sharp phase transition
where the envelope of the local magnetization vanishes uniformly. The
first-order transition in a quantum ferromagnet also remains sharp. The
universal mechanism leading to a tricritical point in an itinerant quantum
ferromagnet is suppressed, and in principle one can recover a quantum critical
point with mean-field exponents. Observable consequences of these results are
discussed.
%
\end{abstract}


\maketitle

In standard phase transitions, such as the paramagnet-ferromagnet transition,
or the liquid-gas transition, a homogeneous order parameter (OP; the
magnetization in a magnet, or the density difference in a fluid) goes to zero
as one crosses from the ordered phase into the disordered one. The OP may
vanish continuously, as in the case of a magnet where the transition is second
order, or discontinuously, as in the case of a fluid where the transition is
first order except at the critical point. An external field may preclude a
homogeneous OP. This happens for a fluid in a gravitational field, which
produces a position-dependent density profile \cite{Sengers_van_Leeuwen_1982}
and in some sense destroys the critical point (see below). Due to the weakness
of gravity, this is a very small effect. This raises the question whether
qualitatively similar, and maybe quantitatively larger, effects can be achieved
in other systems if an external field induces an inhomogeneous OP.

We consider one such example, namely, a ferromagnet subject to mechanical
stress. We will first discuss a classical Heisenberg magnet, and later
generalize to quantum ferromagnets (FMs). Consider a metallic FM in the shape
of a circular disk that is bent in the direction perpendicular to the disk
plane. This leads to a position dependent mass density
\cite{Landau_Lifshitz_VII_1986} and hence, in a metal, to a position dependent
electron density and an inhomogeneous chemical potential $\mu$. The FM
transition is described by a $\phi^4$-theory \cite{Ma_1976}, and within a
Stoner model the inhomogeneous $\mu$ leads to a spatially dependent distance
from criticality. Naively, this means that the system can be tuned to
criticality only at special positions within the sample, not everywhere at the
same time. One might thus expect the transition to become smeared. This is what
appears to be found for the liquid-gas transition in a gravitational field
\cite{Hohenberg_Barmatz_1972, Ahlers_1991, van_Leeuwen_Sengers_1984}, which
does, however, present a physically different situation
\cite{broken_symmetry_footnote}. For the FM case we find that the transition
remains sharp in a well-defined sense with mean-field critical behavior, even
though the magnetization ${\bm M}({\bf x})$ is position dependent and hence
``smeared'' in some sense \cite{Helium_footnote, H(x)_footnote}. One might
expect that ${\bm M}({\bm x})$ is essentially restricted to a surface layer of
fixed width, so that the dimensionality of the system is effectively reduced by
one. This is not the case; we find (for a particular model of a sample of
linear dimension $L$) that ${\bm M}({\bf x})$ is essentially nonzero in a
region of width $L^{1/3}$, so the support of the magnetization diverges in the
thermodynamic limit, $L \to \infty$, albeit more slowly than $L$. This leads to
unusual critical exponents for some spatially averaged observables.

Consider a flat circular disk sample of a metallic FM with thickness $L$ (in
the $x$-direction) and a radius that is some fixed multiple of $L$. A
distortion of the disk in $x$-direction from a flat shape into a paraboloid
leads to a strain tensor whose trace is a linear function of $x$
\cite{Landau_Lifshitz_VII_1986}, and hence to an electron density $n(x) = n_0 +
{\rm const.}\times x$. (This distortion must be achieved by bending, not, e.g.,
by grinding.) Within a Stoner model, the distance $r$ from criticality depends
linearly on the density of states, and hence on the cube root of $n(x)$. We
consider a more general model where $r$ varies as a power of $x$:
\be
r(x) = r_0 + 2(x/L)^n.
\label{eq:1}
\ee
Note that we take the prefactor of the $x$-dependent term to be of $O(1)$ in
order to demonstrate the qualitative effects of such a term. For a real bent
plate the prefactor will be smaller, and we will give a semi-quantitative
discussion below. Also note that $r(L)$ is bounded as $L \to \infty$. This
reflects a bending displacement proportional to $L$ and ensures that a
meaningful thermodynamic limit can be taken. Our model action is a
$\phi^4$-theory with a spatially dependent mass given by $r(x)$,
\be
S = \int_V d{\bm x}\ \left[\frac{r(x)}{2}\,{\bm\phi}^2({\bm x}) +
\frac{c}{2}\left({\bm\nabla}{\bm\phi}({\bm x})\right)^2 +
\frac{u}{4}\left({\bm\phi}^2({\bm x})\right)^2\right].
\label{eq:2}
\ee
The integration extends over a volume $V \propto L^3$, and $c$ and $u$ are
constants. We emphasize that this model is rather general, and a magnet under
stress is only one possible realization. We first treat this problem in a
saddle-point approximation and look for solutions of the form ${\bm\phi}({\bm
x}) = (0,0,M(x))$, with $M(x)$ the inhomogeneous magnetization. The
saddle-point equation then reads
\be
c\,M''(x) = r(x)\,M(x) + u\,M^3(x) - H,
\label{eq:3}
\ee
where we have added a magnetic field $H$. The physical solution of the ODE
(\ref{eq:3}) must obey the boundary conditions $M'(0) = M'(L) = 0$.

While this ODE would be difficult to solve in closed form, we can obtain a
great deal of information from asymptotic solutions and scaling considerations.
For $x\ll L$ one can neglect the $x$-dependence of $r(x)$ and finds a solution
in terms of the Jacobi integral ${\rm sn}(x)$; for $x \approx L$ one can drop
the $M^3$ term and finds a solution in terms of Airy functions. A smeared
transition would imply a nonzero magnetization for all parameter values
\cite{broken_symmetry_footnote}. This is physically not possible: for $r_0>0$,
$r(x)$ is positive definite and the physical solution must be $M(x)\equiv 0$.
We thus expect a sharp phase transition in the following sense: there exists a
value $r_0^c$ of $r_0$ such that the envelope of the magnetization vanishes
uniformly as $r_0 \to r_0^c$ at $H=0$, or as $H \to 0$ at $r_0 = r_0^c$.

There are two explicit length scales in this problem: the zero of $r(x)$, $x_0
= L(-r_0/2)^{1/n}$, and the bare correlation length $\xi_0 = \sqrt{-c/r_0}$.
One expects the phase transition to occur when $x_0 = \xi_0$ (apart from a
factor of $O(1)$). This condition leads to $r_0^c = -2/\ell^{2n/(n+2)}$, with
$\ell = L/\sqrt{c}$ a dimensionless system size. For $r_0 = r_0^c$ the zero of
$r(x)$ is $x_0^c = L/\ell^{2/(n+2)}$. Now define $y = x/x_0^c$, $\mu(y) =
\ell^{n/(n+2)}\,M(x)$, and $h = H\,l^{3n/(n+2)}$. $\mu$ obeys
\bse
\label{eqs:4}
\be
\mu''(y) = \rho(y,r_0\ell^{2n/(n+2)})\,\mu(y) + u\,\mu^3(y) - h,
\label{eq:4a}
\ee
where
\be
\rho(y,z) = z + 2y^n.
\label{eq:4b}
\ee
\ese
The solution of Eq.\ (\ref{eq:4a}) determines $M(x)$ via the relation
\be
M(x) = \ell^{-n/(n+2)}\,\mu\left(x/x_0^c; r_0\,
\ell^{2n/(n+2)},H\,\ell^{3n/(n+2)}\right),
\label{eq:5}
\ee
where we show the dependence of $\mu$ on $r_0$ and $H$.

Now consider the thermodynamic limit, $\ell \to \infty$. Since $r(x)$ is
bounded for all $x$ (see Eq.\ (\ref{eq:1})), we expect physical quantities in
this limit to be independent of $\ell$. From Eq.\ (\ref{eq:5}), this
requirement yields for the local magnetization a power-law prefactor times an
envelope function,
\bse
\label{eqs:6}
\bea
M(x;r_0,H=0) &=& r_0^{1/2}\,g_M^{r_0}(x/x_0^c),
\label{eq:6a}\\
M(x;r_0=0,H) &=& H^{1/3}\,g_M^H(x/x_0^c).
\label{eq:6b}
\eea
Similarly, for the envelope susceptibility $\chi(x) = (\partial M(x)/\partial
H)_{H=0}$ we have
\be
\chi(x;r_0) = r_0^{-1} g_{\chi}(x/x_0^c),
\label{eq:6c}
\ee
\ese
where $g_M^{r_0}$, $g_M^H$, and $g_{\chi}$ are scaling functions. We can
further define a local specific heat $C(x) = \partial^2 f(x)/\partial r_0^2$,
where $f(x)$ is the free energy density which scales as $M^4(x)$. For the
critical exponents $\beta$, $\delta$, $\gamma$, and $\alpha$ defined by $M(x)
\propto r_0^{\beta}$, $M(x) \propto H^{1/\delta}$, $\chi(x) \propto
r_0^{-\gamma}$, and $C(x) \propto r_0^{-\alpha}$ this implies
\bse
\label{eqs:7}
\be
\beta = 1/2 \quad,\quad \delta = 3 \quad,\quad \gamma = 1 \quad,\quad \alpha = 0.
\label{eq:7a}
\ee
Finally, we determine the exponents $\nu$ and $\eta$. The magnetization depends
on $r_0$ only through the combination $r_0\,\ell^{2n/(n+2)}$. If we identify
the diverging length scale $\xi$ that characterizes the phase transition with
$c^{1/2}\ell^{n/(n+2)}$ we have $r_0 \sim \xi^{-2}$, or $\xi \sim r_0^{-\nu}$
with $\nu = 1/2$. Furthermore, the inverse susceptibility determines the
exponent $\eta$ via $\chi^{-1} \propto r_0 \sim \xi^{-2} \equiv \xi^{-2 +
\eta}$ with $\eta = 0$. We thus have
\be
\nu = 1/2 \quad,\quad \eta = 0.
\label{eq:7b}
\ee
\ese
$\nu$ and $\eta$ defined in this way are finite-size scaling exponents, and
$\nu$ does not represent the divergence of a coherence length defined via the
spatial decay of the two-point correlation function. The latter remains finite
even at the transition, as it does in the case of the liquid-gas transition in
a gravitational field \cite{van_Leeuwen_Sengers_1984}. There thus is also a
correlation length exponent $\nu$ that is equal to zero. There is a further
ambiguity within the framework of finite-size scaling: defining $\xi$, for
instance, as $\xi = c^{1/2}\ell$ would lead to different values of $\nu$ and
$\eta$. The mean-field values given above result from what in some sense is the
most natural choice for $\xi$. However, all choices for $\xi$ preserve the
exponent relation $\nu (2-\eta) = \gamma$. The Essam-Fisher relation, $\alpha +
2\beta + \gamma = 2$, is also fulfilled.

\begin{figure}[t,b]
\includegraphics[width=6.0cm]{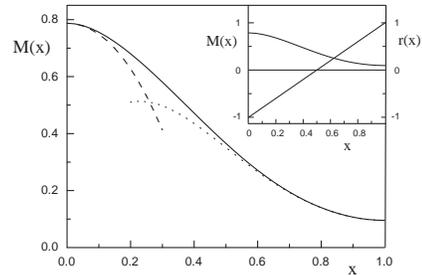}
\vskip -3mm
\caption{The $x$-dependent magnetization (solid line) for $r_0=-1$, $n=1$,
 $L=1$, $c=1/25$, $u=1$, $H=0$. The dashed and dotted lines represent asymptotic
 analytic solutions for $x\ll L$ and $x\approx L$, respectively. The inset shows
 the magnetization together with the $x$-dependent mass function $r(x)$.}
\label{fig:1}
\end{figure}

We have corroborated and augmented these results by solving Eq.\ (\ref{eq:3})
numerically with the boundary condition $M'(0)=0$, and $M(0)$ chosen such that
the mean-field free energy is minimized. Fig.\ \ref{fig:1} shows
$M(x)$ for the case of a linearly $x$-dependent mass, $n=1$. $M(x)$ is large in
the region where $r(x) < 0$ and small in the region where $r(x) > 0$, as one
would expect. The asymptotic solutions mentioned above are also shown in Fig.\
\ref{fig:1}.

\begin{figure}[b]
\includegraphics[width=6.0cm]{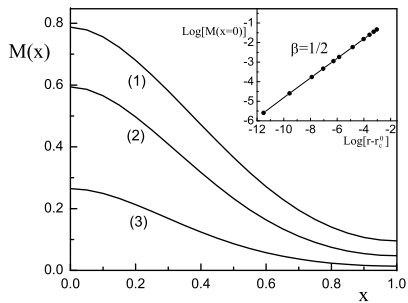}
\vskip -3mm
\caption{The $x$-dependent magnetization for $r_0 = -1$ (1), $r_0 = -0.8$ (2),
 and $r_0 = -0.6$ (3). All other
 parameter values are the same as in Fig.\ \ref{fig:1}, which lead to
 $r_0^c = -0.55213$. The inset shows a log-log
 plot of $M(x=0)$ vs. $r_0 - r_0^c$, with the solid line representing a power law
 with an exponent of $1/2$. See the text for additional information.}
\label{fig:2}
\end{figure}

\begin{figure}[t,h]
\includegraphics[width=6.0cm]{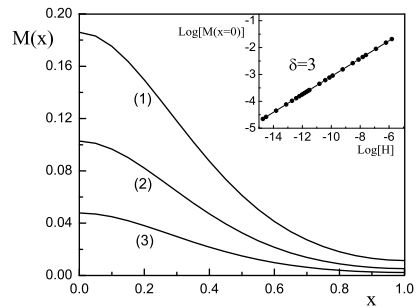}
\vskip -3mm
\caption{The $x$-dependent magnetization for $r_0 = r_0^c$ and $H = 0.003$ (1),
 $H = 0.0005$ (2), and $H = 0.00005$ (3). All other
 parameter values are the same as in Fig.\ \ref{fig:1}.
 The inset shows a log-log plot of
 $M(x=0)$ vs $H$, with the solid line
 representing a power law with an exponent of $1/3$.}
\label{fig:3}
\end{figure}

With increasing $r_0$, the magnetization decreases for all $x$, see Fig.\
\ref{fig:2}, and it vanishes uniformly when $r_0$ reaches the critical value
$r_0^c$. For $r_0 > r_0^c$ the physical solution of Eq.\ (\ref{eq:3}) is $M(x)
\equiv 0$. As $r_0 \to r_0^c$ from below, the envelope of the magnetization
vanishes as $\vert r_0 - r_0^c\vert^{1/2}$. For the value $M(x=0)$ this is
demonstrated in the inset in Fig.\ \ref{fig:2}. In a magnetic field the
magnetization vanishes uniformly as $H^{1/3}$ for $H \to 0$ at $r_0 = r_0^c$,
see Fig.\ \ref{fig:3}.
These results show that there is a sharp phase transition: the envelope of the
magnetization vanishes uniformly as $r_0 \to r_0^c$ at $H=0$, or as $H \to 0$
at $r_0 = r_0^c$. The order-parameter critical exponents have mean-field
values: $\beta = 1/2$ and $\delta = 3$. The numerics suggest that the
mean-field critical behavior also holds for finite $L$. We have found Eqs.\
(\ref{eq:6a}, \ref{eq:6b}) to hold for values of $L^2/c$ as small as 16.

We now consider spatially averaged observables (denoted by an overbar) rather
than local quantities. Consider an averaged magnetization ${\bar M} =
(1/L)\int_0^L dx\ M(x)$. Since $M(x)$ is essentially nonzero only on the
interval $x \in [0,x_0^c]$, the upper limit of the integral is essentially
$x_0^c \propto \ell^{n/(n+2)}$. Corrections to this approximation are
exponentially small. ${\bar M}$ obeys, instead of Eq.\ (\ref{eq:5}),
\bse
\label{eqs:8}
\bea
{\bar M} &=&
\ell^{-1}\,f_M\left(r_0\,\ell^{2n/(n+2)},H\,\ell^{3n/(n+2)}\right),
\label{eq:8a}\\
f_M(u,v) &=& \int_0^1 dy\,\mu(y;u,v).
\label{eq:8b}
\eea
\ese
Demanding again that observables are independent of $\ell$ for $\ell \to
\infty$, this leads to exponents ${\bar\beta} = (n+2)/2n$ and ${\bar\delta} =
3n/(n+2)$. Analogous considerations for the other observables we have
considered yield the following set of exponents for spatially averaged
quantities,
\bse
\label{eqs:9}
\be
{\bar\beta} = \frac{n+2}{2n} \ ,\ {\bar\delta} = \frac{3n}{n+2} \ ,\
{\bar\gamma} = \frac{n-1}{n} \ ,\  {\bar\alpha} = \frac{-1}{n} \ .
\label{eq:9a}
\ee
From the averaged susceptibility ${\bar\chi} = (\partial{\bar M}/\partial
H)_{H=0}$ we find the exponent ${\bar\eta}$, and $\nu$ is unchanged since
${\bar M}$ depends on the same combination of $r_0$ and $\ell$ as $M(x)$,
\be
{\bar\eta} = 2/n \quad,\quad {\bar\nu} = 1/2.
\label{eq:9b}
\ee
\ese
These exponents for the averaged quantities satisfy again the relations
${\bar\alpha} + 2{\bar\beta} + {\bar\gamma} = 2$ and ${\bar\nu}(2-{\bar\eta}) =
{\bar\gamma}$.

The exponent values depend on how the averaged quantities are defined, and
hence on what exactly is being measured. For instance, if one defined ${\bar
M}$ as $(1/c^{1/3}L^{1/3}) \int_0^L dx\ M(x)$ to account for the fact that the
magnetization is essentially nonzero only for $x < x_0^c \propto \ell^{1/3}$,
one would find mean-field values for all exponents.

We now apply these results to quantum FMs. It is observed that the transition
in itinerant FMs at sufficiently low temperatures $T$ is always first order
\cite{Uemura_et_al_2007}. This has been explained in terms of fluctuation
effects due to the coupling of the OP fluctuations to soft particle-hole
excitations in metallic FMs. As a result, the free energy in a mean-field
approximation has the form \cite{Belitz_Kirkpatrick_Vojta_1999}
\be
F = \frac{r}{2}\,M^2 + \frac{v}{4}\,M^4\ln(M^2 + T^2) + \frac{u}{4}\,M^4 +
O(M^6),
\label{eq:10}
\ee
with $v>0$. The $v$-term is negative, which leads to a first-order transition
at $T=0$, and to a tricritical point at a temperature $T_{\rm tc} =
\exp(-u/2v)$ \cite{compressibility_footnote}.

A position dependent chemical potential in the regime where the transition is
first order has two effects: (1) It lowers the tricritical temperature $T_{\rm
tc}$, and a sufficiently strong space dependence restores a quantum critical
point; (2) it leads to an inhomogeneous magnetization as in the case of
classical magnets discussed above.

To quantify the first effect we estimate the prefactor of the space dependent
term in $\mu$. For a displacement at the edge of the plate equal to $0.01\,L$
(conservatively; more severe bending may be possible) one finds a density
variation of about $0.01\,n_0$, and, instead of Eq.\ (\ref{eq:1}), $r(x) = r_0
+ 2\rho\,x/L$ with $\rho = O(0.01)$. Suppressing factors of $O(1)$, the
chemical potential is $\mu(x) = \epsilon_{\rm F}(1 + \rho\,x/L)$. The effects
are smaller by a factor of about $10^2$ compared to those shown in our figures,
but still large compared to those of gravity on a fluid. To estimate the effect
on $T_{\rm tc}$ we recall that the $v$-term in the free energy is due to a soft
propagator $P = 1/(\omega + k v_{\text{F}})$
\cite{Belitz_Kirkpatrick_Vojta_2005}, with $\omega$ ($k$) the frequency (wave
number). A position dependent $\mu$ causes $P$ to acquire a mass or cutoff
frequency $\omega_{\text{c}}$ proportional to the prefactor of the
$x$-dependency in $\mu(x)$, $\omega_{\text{c}} \approx
\rho\,\epsilon_{\text{F}}/\hbar L k_{\text{F}}$. $T_{\text{tc}}$ is lowered by
roughly $\hbar \omega_c / k_{\text{B}}$. With $\rho$ as above, and typical
values for all other quantities, one finds a very small suppression $\delta
T_{\text{tc}} = O(10^{-6}\,\text{K})$. It is interesting, however, that a
sufficiently strong spatial dependence of $\mu$ will eventually destroy the
mechanism that leads to a first-order quantum phase transition, as does
quenched disorder \cite{Belitz_Kirkpatrick_Vojta_2005}.

\begin{figure}[t]
\includegraphics[width=6.0cm]{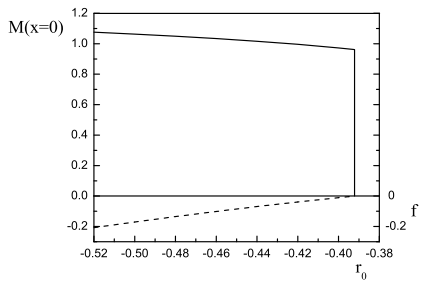}
\vskip -3mm
\caption{$M(x=0)$ vs. $r_0$ (solid line, left scale) as obtained from the
 solution of Eq.\ \ref{eq:11} for parameter values $L=1$, $c=1/25$, $u=-1$,
 $w = 1$, which lead to $r_0^1 = -0.39216$. Also shown is the free energy
 density $f$ (dashed line, right scale).}
\label{fig:4}
\end{figure}

To study the second effect we thus assume that the transition is still first
order at low temperature. In this case the logarithm in Eq.\ (\ref{eq:10}) can
be expanded, and the Landau free energy is adequately represented by a power
series with a negative $M^4$ term and a positive $M^6$ term. Generalizing to a
non-homogeneous situation we then obtain the following ODE for the OP in the
quantum case,
\be
c\,M''(x) = r(x)\,M(x) + u\,M^3(x) + w\,M^5(x),
\label{eq:11}
\ee
which replaces Eq.\ (\ref{eq:3}). Here $w>0$, $u<0$, and $r(x)$ is given by
Eq.\ (\ref{eq:1}). The resulting $M(x)$ curves are very similar to those in
Fig.\ \ref{fig:2}, but now $M(x)$ goes to zero discontinuously at a critical
value $r_0^1$ of $r_0$, see Fig.\ \ref{fig:4}.

We now consider the validity of our mean-field treatment. If the quantum
critical point is restored by a (hypothetical) large position dependence of
$\mu$, the quantum critical behavior will be mean-field like since the spatial
dependence of $\mu$ suppresses the mechanism that causes a first-order
transition in clean quantum FMs, and non-mean-field critical behavior in
disordered ones \cite{Belitz_Kirkpatrick_Vojta_2005}. For the classical
transition it is possible that the mean-field critical behavior will also be
exact. This is because the two-point correlation function is always finite, and
hence there are no divergencies in simple perturbation theory. However, the
divergence of the envelope susceptibility, Eq.\ (\ref{eq:6c}), shows that there
are fluctuations that may influence the critical behavior. Fluctuations of the
elastic deformation should also be considered\cite{Larkin_Pikin_1969,
Pikin_1970}. These points require additional investigation.

In summary, we have considered a model for a metallic FM with a position
dependent electron density or chemical potential. This can be realized by
mechanically stressing the sample. The phase transition remains sharp even
though the electron density is not homogeneous, and we have given critical
exponents for both local and spatially averaged observables. In the quantum
case, the tricritical temperature is lowered, although for realistic stresses
this is a small effect. If a stronger position dependence can be realized
(e.g., by means of optical lattices), this will eventually suppress the
tricritical point, restoring a quantum critical point with mean-field critical
behavior.

This work was supported by the NSF under grant Nos. DMR-05-29966 and
DMR-05-30314. We would like to thank Jan Sengers and John Toner for useful
discussions.

\vskip -1mm

\begin{thebibliography}{16}
\expandafter\ifx\csname natexlab\endcsname\relax\def\natexlab#1{#1}\fi
\expandafter\ifx\csname bibnamefont\endcsname\relax
  \def\bibnamefont#1{#1}\fi
\expandafter\ifx\csname bibfnamefont\endcsname\relax
  \def\bibfnamefont#1{#1}\fi
\expandafter\ifx\csname citenamefont\endcsname\relax
  \def\citenamefont#1{#1}\fi
\expandafter\ifx\csname url\endcsname\relax
  \def\url#1{\texttt{#1}}\fi
\expandafter\ifx\csname urlprefix\endcsname\relax\def\urlprefix{URL }\fi
\providecommand{\bibinfo}[2]{#2} \providecommand{\eprint}[2][]{\url{#2}}

\bibitem[{\citenamefont{Sengers and van
  Leeuwen}(1982)}]{Sengers_van_Leeuwen_1982}
\bibinfo{author}{\bibfnamefont{J.~V.} \bibnamefont{Sengers}} \bibnamefont{and}
  \bibinfo{author}{\bibfnamefont{J.~M.~J.} \bibnamefont{van Leeuwen}},
  \bibinfo{journal}{Physica} \textbf{\bibinfo{volume}{116A}},
  \bibinfo{pages}{345} (\bibinfo{year}{1982}).

\bibitem[{\citenamefont{Landau and Lifshitz}(1986)}]{Landau_Lifshitz_VII_1986}
\bibinfo{author}{\bibfnamefont{L.~D.} \bibnamefont{Landau}} \bibnamefont{and}
  \bibinfo{author}{\bibfnamefont{E.~M.} \bibnamefont{Lifshitz}},
  \emph{\bibinfo{title}{Theory of Elasticity}} (\bibinfo{publisher}{Pergamon,
  Oxford}, \bibinfo{year}{1986}).

\bibitem[{\citenamefont{Ma}(1976)}]{Ma_1976}
\bibinfo{author}{\bibfnamefont{S.-K.} \bibnamefont{Ma}},
  \emph{\bibinfo{title}{Modern Theory of Critical Phenomena}}
  (\bibinfo{publisher}{Benjamin, Reading, MA}, \bibinfo{year}{1976}).

\bibitem[{\citenamefont{Hohenberg and Barmatz}(1972)}]{Hohenberg_Barmatz_1972}
\bibinfo{author}{\bibfnamefont{P.~C.} \bibnamefont{Hohenberg}}
  \bibnamefont{and} \bibinfo{author}{\bibfnamefont{M.}~\bibnamefont{Barmatz}},
  \bibinfo{journal}{Phys. Rev. A} \textbf{\bibinfo{volume}{6}},
  \bibinfo{pages}{289} (\bibinfo{year}{1972}).

\bibitem[{\citenamefont{Ahlers}(1991)}]{Ahlers_1991}
\bibinfo{author}{\bibfnamefont{G.}~\bibnamefont{Ahlers}}, \bibinfo{journal}{J.
  Low Temp. Phys.} \textbf{\bibinfo{volume}{84}}, \bibinfo{pages}{173}
  (\bibinfo{year}{1991}).

\bibitem[{\citenamefont{van Leeuwen and
  Sengers}(1984)}]{van_Leeuwen_Sengers_1984}
\bibinfo{author}{\bibfnamefont{J.~M.~J.} \bibnamefont{van Leeuwen}}
  \bibnamefont{and} \bibinfo{author}{\bibfnamefont{J.~V.}
  \bibnamefont{Sengers}}, \bibinfo{journal}{Physica}
  \textbf{\bibinfo{volume}{128A}}, \bibinfo{pages}{99} (\bibinfo{year}{1984}).

\bibitem[{bro()}]{broken_symmetry_footnote}
\bibinfo{note}{In the magnetic case there is a broken symmetry, whereas in the
  liquid-gas case there is not. A ``smeared transition'' implies an OP that is
  nonzero everywhere, albeit very small in some regions of parameter space. If
  a symmetry is broken, this implies that it is always broken.}

\bibitem[{Hel()}]{Helium_footnote}
\bibinfo{note}{The $\lambda$-transition in He 4 is described by an XY-model
  with a spatially dependent mass. The prevailing view is that in this case,
  too, there is no sharp transition \cite{Ahlers_1991}. However, the actual
  observations are consistent with a sharp phase transition in the sense
  described here.}

\bibitem[{H(x()}]{H(x)_footnote}
\bibinfo{note}{A fluid in a gravitational field is more closely analogous to a
  magnet in a position-dependent magnetic field, which would destroy the
  transition.}

\bibitem[{\citenamefont{Uemura and et~al.}(2007)}]{Uemura_et_al_2007}
\bibinfo{author}{\bibfnamefont{Y.~J.} \bibnamefont{Uemura}} \bibnamefont{and}
  \bibinfo{author}{\bibnamefont{et~al.}}, \bibinfo{journal}{Nature Physics}
  \textbf{\bibinfo{volume}{3}}, \bibinfo{pages}{29} (\bibinfo{year}{2007}),
  \bibinfo{note}{and references therein}.

\bibitem[{\citenamefont{Belitz et~al.}(1999)\citenamefont{Belitz, Kirkpatrick,
  and Vojta}}]{Belitz_Kirkpatrick_Vojta_1999}
\bibinfo{author}{\bibfnamefont{D.}~\bibnamefont{Belitz}},
  \bibinfo{author}{\bibfnamefont{T.~R.} \bibnamefont{Kirkpatrick}},
  \bibnamefont{and} \bibinfo{author}{\bibfnamefont{T.}~\bibnamefont{Vojta}},
  \bibinfo{journal}{Phys. Rev. Lett.} \textbf{\bibinfo{volume}{82}},
  \bibinfo{pages}{4707} (\bibinfo{year}{1999}).

\bibitem[{com()}]{compressibility_footnote}
\bibinfo{note}{In a compressible magnet the transition can be first order due
  to coupling to phonons \cite{Larkin_Pikin_1969, Bergman_Halperin_1976}; see
  also Ref.\ \onlinecite{Pikin_1970}, which included some fermionic fluctuations.
  For our Heisenberg model this happens only if the magnetoelasic coupling
  $\lambda$ exceeds a critical strength $\lambda_{\text c}$. We assume $\lambda
  < \lambda_{\text c}$, or that the transition resulting from this mechanism is
  very weakly first order.}

\bibitem[{\citenamefont{Belitz et~al.}(2005)\citenamefont{Belitz, Kirkpatrick,
  and Vojta}}]{Belitz_Kirkpatrick_Vojta_2005}
\bibinfo{author}{\bibfnamefont{D.}~\bibnamefont{Belitz}},
  \bibinfo{author}{\bibfnamefont{T.~R.} \bibnamefont{Kirkpatrick}},
  \bibnamefont{and} \bibinfo{author}{\bibfnamefont{T.}~\bibnamefont{Vojta}},
  \bibinfo{journal}{Rev. Mod. Phys.} \textbf{\bibinfo{volume}{77}},
  \bibinfo{pages}{579} (\bibinfo{year}{2005}).

\bibitem[{\citenamefont{Pikin}(1970)}]{Pikin_1970}
\bibinfo{author}{\bibfnamefont{S.~A.} \bibnamefont{Pikin}},
  \bibinfo{journal}{Zh. Eksp. Teor. Fiz. {\bf 58}, 1406}
  (\bibinfo{year}{1970}), \bibinfo{note}{[Sov. Phys. JETP {\bf 31}, 753
  (1970)]}.

\bibitem[{\citenamefont{Larkin and Pikin}(1969)}]{Larkin_Pikin_1969}
\bibinfo{author}{\bibfnamefont{A.~I.} \bibnamefont{Larkin}} \bibnamefont{and}
  \bibinfo{author}{\bibfnamefont{S.~A.} \bibnamefont{Pikin}},
  \bibinfo{journal}{Zh. Eksp. Teor. Fiz.} \textbf{\bibinfo{volume}{56}},
  \bibinfo{pages}{1664} (\bibinfo{year}{1969}), \bibinfo{note}{[Sov. Phys. JETP
  {\bf 29}, 891 (1969)]}.

\bibitem[{\citenamefont{Bergman and Halperin}(1976)}]{Bergman_Halperin_1976}
\bibinfo{author}{\bibfnamefont{D.~J.} \bibnamefont{Bergman}} \bibnamefont{and}
  \bibinfo{author}{\bibfnamefont{B.~I.} \bibnamefont{Halperin}},
  \bibinfo{journal}{Phys. Rev. B} \textbf{\bibinfo{volume}{13}},
  \bibinfo{pages}{2145} (\bibinfo{year}{1976}), \bibinfo{note}{and references
  therein}.

\end{thebibliography}

\end{document}